\documentclass[preprint,showpacs,preprintnumbers,amsmath,amssymb]{revtex4}

\usepackage{graphicx}
\usepackage{dcolumn}
\usepackage{bm}

\begin{document}
\preprint{}

\title{Pressure-induced phase transition of Bi$_2$Te$_3$ into the bcc structure}

\author{Mari Einaga}%
  \email{einaga.mari@phys.sc.niigata-u.ac.jp}
 \affiliation{Graduate School of science \& Technology, Niigata University 8050 Ikarashi-Ninocho, Nishi-ku, Niigata, Niigata 950-2181, Japan}
\author{Ayako Ohmura}
\affiliation{Center for Transdisciplinary Research, Niigata University 8050 Ikarashi-Ninocho, Nishi-ku, Niigata, Niigata 950-2181, Japan}
\author{Atsuko Nakayama}
\affiliation{Center for Transdisciplinary Research, Niigata University 8050 Ikarashi-Ninocho, Nishi-ku, Niigata, Niigata 950-2181, Japan}
\author{Fumihiro Ishikawa}
 \affiliation{Graduate School of science \& Technology, Niigata University 8050 Ikarashi-Ninocho, Nishi-ku, Niigata, Niigata 950-2181, Japan}
\author{Yuh Yamada}
\affiliation{Department of Physics, Niigata University 8050 Ikarashi-Ninocho, Nishi-ku, Niigata, Niigata 950-2181, Japan}%

\author{Satoshi Nakano}
\affiliation{National Institute for Materials Science (NIMS), 1-1 Namiki, Tsukuba 305-0044, Japan}%

\date{\today}

\begin{abstract}
The pressure-induced phase transition of bismuth telluride, Bi$_2$Te$_3$, has been studied by synchrotron x-ray diffraction measurements at room temperature using a diamond-anvil cell (DAC) with loading pressures up to 29.8 GPa. We found a high-pressure body-centered cubic (bcc) phase in Bi$_2$Te$_3$ at 25.2 GPa, which is denoted as phase IV, and this phase apperars above 14.5 GPa. Upon releasing the pressure from 29.8 GPa, the diffraction pattern changes with pressure hysteresis. The original rhombohedral phase is recovered at 2.43 GPa. The bcc structure can explain the phase IV peaks. We assumed that the structural model of phase IV is analogous to a substitutional binary alloy; the Bi and Te atoms are distributed in the bcc-lattice sites with space group $Im\bar{3}m$. The results of Rietveld analysis based on this model agree well with both the experimental data and calculated results. Therefore, the structure of phase IV in Bi$_2$Te$_3$ can be explained by a solid solution with a bcc lattice in the Bi$\--$Te (60 atomic\% tellurium) binary system.
\end{abstract}

\maketitle

\section{\label{sec:level1}Introduction}

Bismuth telluride, Bi$_2$Te$_3$, is a typical thermoelectric material with a high-performance near room temperature \cite{1, 2, 3}. Bi$_2$Te$_3$ has a rhombohedral structure with a space group $R\bar{3}m$, which is denoted as phase I; the hexagonal unit-cell parameters are $a_0$ = 4.395~\AA~and $c_0$ = 30.44~\AA~at ambient pressure and temperature \cite{4}. Bulk Bi$_2$Te$_3$ has a relatively narrow band-gap of 0.171 eV \cite{5} and a high density of states near the Fermi level. Upon applying pressure, these characters are expected to cause metallization and superconductivity \cite{5, 6, 7, 8, 9, 10, 11, 12}. Several groups have reported pressure-induced superconductivity \cite{9, 10, 11, 12, 13, 14}. In our previous study under hydrostatic pressure, the onset temperature of the superconducting transition ${T_{c}}^\mathrm{onset}$ is 2.7 K at 9.0 GPa. As the pressure increases, ${T_{c}}^\mathrm{onset}$ decreases up to 10 GPa, but greatly increases from 10 to 13 GPa: ${T_{c}}^\mathrm{onset}$ = 5.0 K at 13 GPa \cite{12}.

High-pressure x-ray diffraction studies have shown that the structural phase transition from phase I to phase II occurs around 8 GPa \cite{11, 15}. The phase II coexists phase III under pressure above 14 GPa \cite{11}. The crystal structures of phase II and III have yet to be determined. Jacobsen \textit{et} \textit{al.} have performed x-ray diffraction measurements using an ethanol$\--$methanol mixture as transmitting medium, and found the phase II crystal has orthorhombic $I222$ symmetry \cite{15}. Compared to the structural changes under high pressure, the negative and positive pressure dependences of $T_c$ are assumed to be due to phase II and phase III, respectively \cite{12}.

Furthermore, a recent theoretical study predicted Bi$_2$Te$_3$, bismuth selenide Bi$_2$Se$_3$, and bismuth$\--$antimony binary alloy Bi$_{1-x}$Sb$_x$ are candidates for three-dimensional topological insulators \cite{16}, and they have been experimentally established \cite{17}. Since then, Bi$_2$Te$_3$ has attracted much attention in basic and applied research. Herein we conduct x-ray diffraction measurements under hydrostatic pressure up to 30 GPa to reveal the structure at higher pressures because structural information is crucial for understanding various phenomena.

\section{Experimental}

The sample was a lumps of polycrystalline Bi$_2$Te$_3$ (99.99\% purity, Kojundo Chemical Lab. Co., Ltd.), which was cooled by liquid nitrogen in an alumina mortar and subsequently ground into a fine powder over 10 hours under nitrogen gas. The sample was pressurized using a diamond anvil cell (DAC) assembled by a pair of diamond anvils with 0.3 mm culet-diameter and 2 mm anvil-height. The indentation technique using the anvil culet surface reduced the rhenium gasket thickness from 150 $\mu\textrm{m}$ to 45 $\mu\textrm{m}$. The sample chamber was prepared by drilling a 160 $\mu\textrm{m}$ diameter hole at the center of the indentation on the gasket. The powdered sample was placed in the chamber with ruby balls ($\sim$10 $\mu\textrm{m}$ in diameter) as a pressure marker \cite{18}. Then the chamber was filled with high-density helium gas as the pressure-transmitting medium, which was compressed up to 180 MPa at room temperature by a gas loading system \cite{19}.

Angle-dispersive powder x-ray diffraction measurements were carried out in beamline BL-18C of the Photon Factory in High Energy Accelerator Research Organization (KEK), Tsukuba, Japan. The sample in the DAC was irradiated using synchrotron radiation beams monochromatized to an energy of 25.6 keV ($\lambda$ $\approx$ 0.4840 \AA) through a pinhole collimator with a 40 $\mu\textrm{m}$ in diameter. Each diffraction pattern was recorded using an imaging plate (200 mm$\times$250 mm area) with an exposure time between 120$\--$540 min upon compression up to 29.8 GPa and upon decompression at room temperature.

\section{Results and Discussion}

Figure 1 shows the x-ray diffraction patterns of Bi$_2$Te$_3$. All the reflections obtained at 0.61 GPa are explained by space group $R\bar{3}m$ with lattice parameters of $a$ = 4.366(0)~\AA~and $c$ = 30.11(0)~\AA. The reflections from phase II \cite{11} are observed at pressure above 8.41 GPa. These intensities are enhanced under pressure up to about 20 GPa. After the transition from phase I to phase II, we found small reflections, which are marked by $\bigtriangledown$ and $\downarrow$ at pressure above 14.5 GPa in Fig. 1 and differ from the phase II reflections. This mixed-phase state continues to 23.1 GPa. However, the crystal structure transforms to the high-symmetry one at the pressures above 25.2 GPa; a single phase assigned to a cubic system is obtained. From the changes in the patterns, reflections $\bigtriangledown$ and $\downarrow$ are assigned to phase III and phase IV, respectively. As the pressure decreases from 29.8 GPa, the diffraction patterns exhibit the opposite change with a pressure hysteresis. At 10.3 GPa, phase II and III reappear, and the original rhombohedral phase I reappears at 2.43 GPa. As shown in the top of Fig. 1, phase I is completely recovered upon releasing the pressure.

A body-centered cubic (bcc) structure can explain the peaks from phase IV obtained at 25.2 GPa. Except for the bcc-phase peaks, extra reflections are not present [Fig. 2(a)], suggesting the structural model of phase IV is analogous to substitutional binary alloys; the Bi and Te atoms with their original concentrations are arranged in the bcc-sites. For structural refinement, we assumed the structure of phase IV has a probability of Bi 40\% and Te 60\% atoms and the atoms are arranged in the bcc-lattice sites with space group $Im\bar{3}m$. Rietveld refinement of phase IV was performed for the diffraction patterns obtained at 23.1, 25.2, 27.4, and 29.8 GPa using RIETAN-2000 program \cite{20, 21}. In the present analysis, the calculated patterns agree well the experimental ones within $R_\mathrm{wp}$ = 2.54$\--$2.75\%, $S$ = 0.770$\--$1.12 at each pressure. Figure 2(b) shows the result of Rietveld analysis for the diffraction pattern obtained at 25.2 GPa with $R$ factor $R_\mathrm{wp}$ = 2.73\% and $S$ = 0.830, yielding a lattice constant of $a$ = 3.583(0) \AA. Therefore, we determined that the structure of phase IV in Bi$_2$Te$_3$ can be explained by a solid solution with a bcc-lattice in the Bi$\--$Te binary system. Figure 3 shows the pressure change with atomic volume in phase IV. The atomic volume continuously decreases with increasing pressure. At 29.8 GPa, the atomic volume reaches 66.4\% of the original volume at ambient pressure.

We evaluated the homogeneity of the bcc-solid solution in the Bi$\--$Te system through Vegard's law \cite{22, 23}, which is an empirical rule to explain the character of a solid solution. The law holds that a linear relation, which can be explained by a hard sphere model \cite{23}, exists between the lattice constant of an alloy and the concentrations of the constituent elements at a constant temperature \cite{22}. As with Bi$_2$Te$_3$, the crystal structures of pure bismuth and tellurium also exhibit high-pressure bcc-phases \cite{24, 25}. Therefore, we compared the bcc-phase atomic-volume of Bi$_2$Te$_3$ to that of Bi$_{0.4}$Te$_{0.6}$ obtained from a linear interpolation between pure bismuth and tellurium. Each atomic volume of bismuth and tellurium was estimated by the equation of state curve reported in previous studies \cite{24, 25}. At each pressure, we initially estimated the atomic volumes of the bcc-phase in pure bismuth and tellurium by interpolation and extrapolation using first-order Murnaghan and Vinet equations of state, respectively. Then, we estimated the atomic volumes of Bi$_2$Te$_3$ as a solid solution of 60 atomic\% tellurium concentration, Bi$_{0.4}$Te$_{0.6}$. Finally, we compared the atomic volumes of Bi$_2$Te$_3$ to those of pure bismuth, tellurium, and Bi$_{0.4}$Te$_{0.6}$ under high pressure. Figure 4 shows the pressure dependence of the atomic volumes of the high-pressure bcc-phases in experimentally observed Bi$_2$Te$_3$, estimated Bi$_\mathrm{0.4}$Te$_\mathrm{0.6}$, pure bismuth, and pure tellurium. The atomic volumes of Bi$_2$Te$_3$ greatly deviate from the values estimated by Vegard's law $V_\mathrm{Vegard}$ when the deviation is defined as $V_\mathrm{dev} = 1-V_\mathrm{Vegard}/V_\mathrm{exp}$ where $V_\mathrm{exp}$ and $V_\mathrm{Vegard}$ are the experimentally obtained atomic volumes and the estimated one from Vegard's law, respectively. From the definition, the estimated deviation is 3.702$\--$4.873\%, and $V_\mathrm{exp}$ is larger than $V_\mathrm{Vegard}$ at each pressure. At ambient pressure, the lattice constants and tellurium concentration has been compared in a metastable solid solution containing up to 50 atomic\% tellurium by an ultra-high cooling rate of the melt \cite{26}. In the concentrate region, the experimentally-obtained atomic volume is larger than one estimated from the hard-sphere model. Meanwhile, the bismuth$\--$antimony alloy Bi$_{1-x}$Sb$_x$, which is a typical binary complete solid solution, shows a linear behavior with Vegard's law at ambient conditions. Additionally, Bi$_{1-x}$Sb$_x$ forms the bcc structure under high pressure; the phase transition to the bcc structure occurs at 12 GPa for $x$ = 0.15 \cite{27, 28}. Furthermore, we have clearly shown the atomic volume of Bi$_{0.85}$Sb$_{0.15}$ follows Vegard's law under high pressure \cite{27}.

At ambient pressure, Bi$_{1-x}$Sb$_x$, which is composed of elements in the 15th group of the periodic table, forms covalent bonds \cite{29}. On the other hand, in Bi$_2$Te$_3$, which is composed of elements in the 15th and 16th groups, the covalent bonds have an ionic component along Bi$\--$Te bonds, which causes the bonds to be stronger than normal covalent bonds \cite{30}. Therefore, we infer that the deviation from Vegard's law in phase IV of Bi$_2$Te$_3$ is due to the remaining strong ionic-covalent bonds in the pressure region of phase IV. Upon further compression, the ionic-covalent bonds should disappear, and the atomic volume of Bi$_2$Te$_3$ will obey Vegard's law.

\section{Summary}

Synchrotron-radiation x-ray diffraction measurements were used to investigate the high-pressure structural phase of Bi$_2$Te$_3$ up to 29.8 GPa. We found a high-pressure bcc-phase in Bi$_2$Te$_3$ at 25.2 GPa. The structure of phase IV is explained by a solid solution with a bcc-lattice in space group $Im\bar{3}m$ for a Bi$\--$Te binary system; the Bi and Te atoms are distributed in the bcc-lattice sites. Due to the remaining strong ionic-covalent bonds up to 29.8 GPa, the relation between the atomic volume and concentration of the constituent elements deviates from Vegard's law.

\begin{acknowledgments}
The present work was concluded under Proposal No. 2009G656 of the Photon Factory. This research was partially supported by Grant-in-Aids for Science Research on Priority Areas: Novel State of Matter Induced by Frustration No. 20046006, and New Materials Science Using Regulated Nano Space-Strategy in Ubiquitous Elements, No. 20045003, from the Ministry of Education, Culture, Sports, Science, and Technology of Japan.
\end{acknowledgments}

\clearpage

\begin{figure}
\begin{center}
\includegraphics[width=15cm,bb=0 0 609 815]{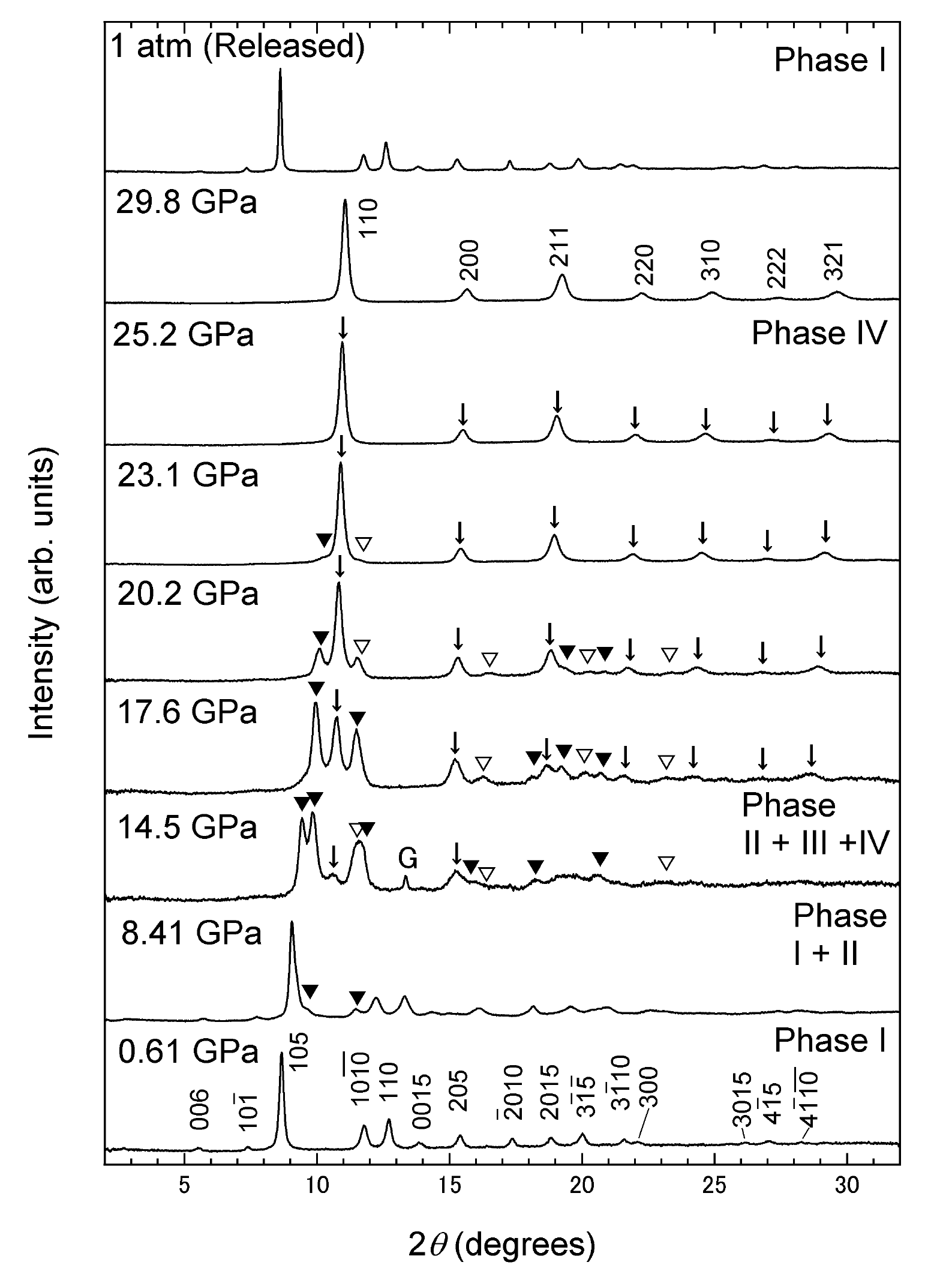}
\caption{\label{fig:AS and CP3} X-ray diffraction patterns of Bi$_2$Te$_3$ under pressure up to 29.8 GPa at room temperature. Arrows indicate reflection of phase IV. Closed and open triangles indicate reflections of phases II and III, respectively. Top pattern was observed at ambient pressure after releasing pressure. Letter G denotes reflection from gasket.}
\end{center}
\end{figure}

\clearpage

\begin{figure}
\includegraphics[width=12cm,bb=0 0 314 156]{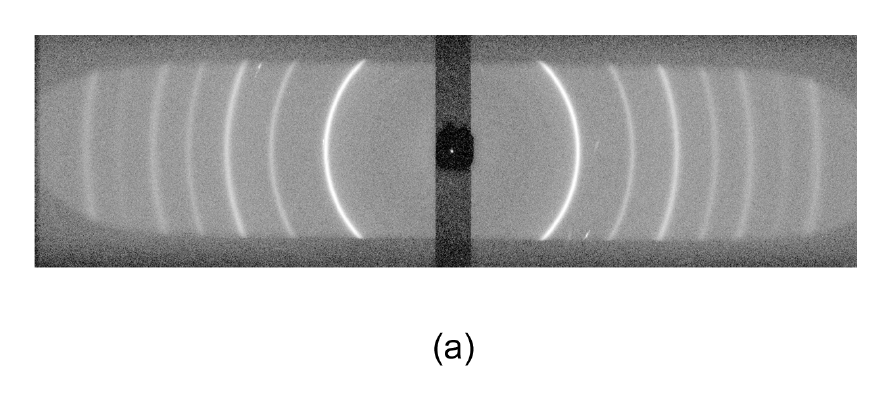}

\includegraphics[width=15cm,bb= 0 0 624 493]{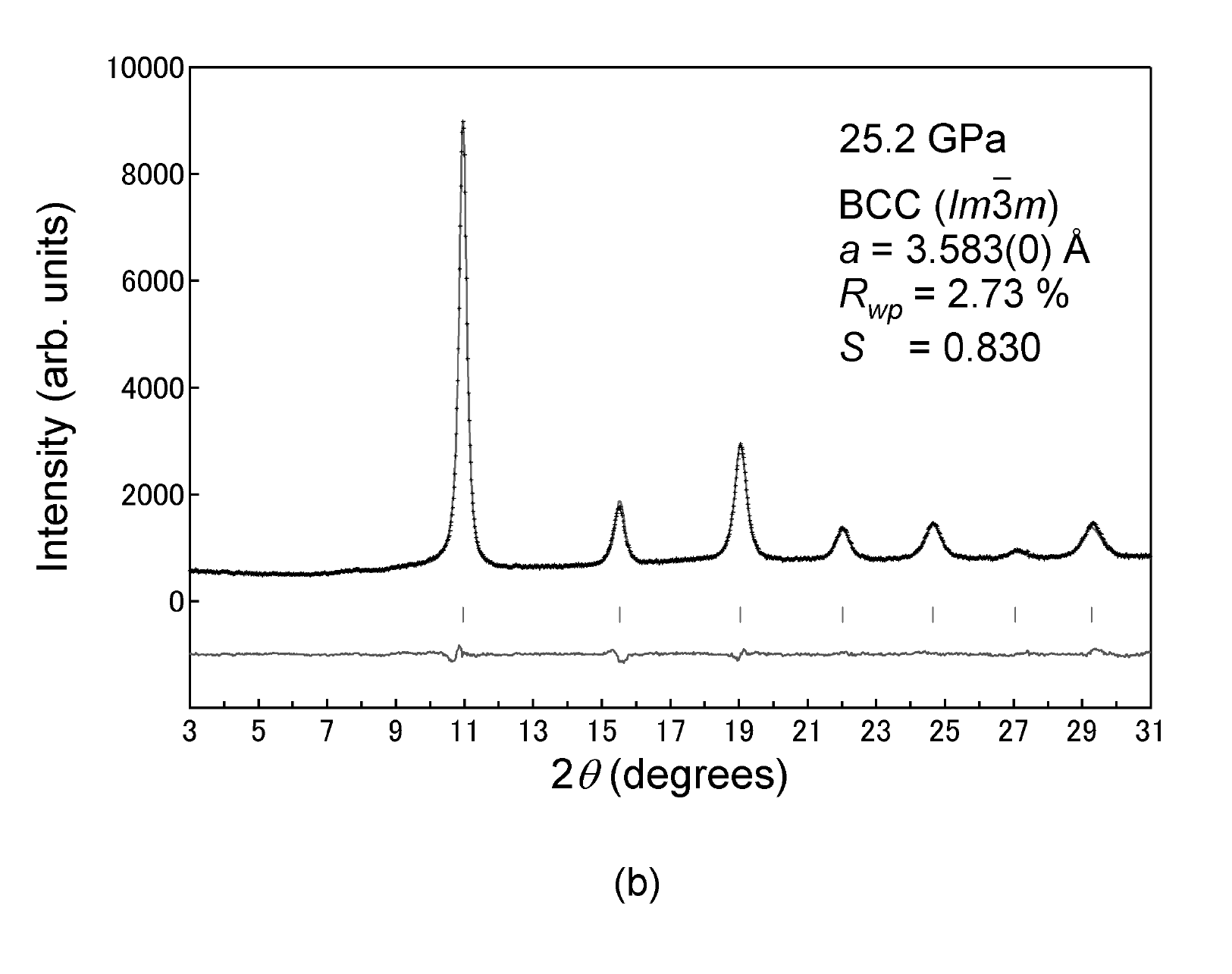}
\caption{\label{fig:S(Q) on compression} (a) Powder diffraction images of Bi$_2$Te$_3$ at 25.2 GPa recorded on an imaging plate. The spot in the diffraction image come from Kossel lines of diamond anvil, which were removed by image processing. (b) Result of Rietveld analysis of Bi$_2$Te$_3$ at 25.2 GPa. Dots and solid line represent the observed and calculated intensities, respectively. Ticks below the profile mark the positions of the reflections from the bcc lattice. Solid line at the bottom shows the residual error.}
\end{figure}

\clearpage

\begin{figure}
\includegraphics[width=15cm,bb=0 0 557 371]{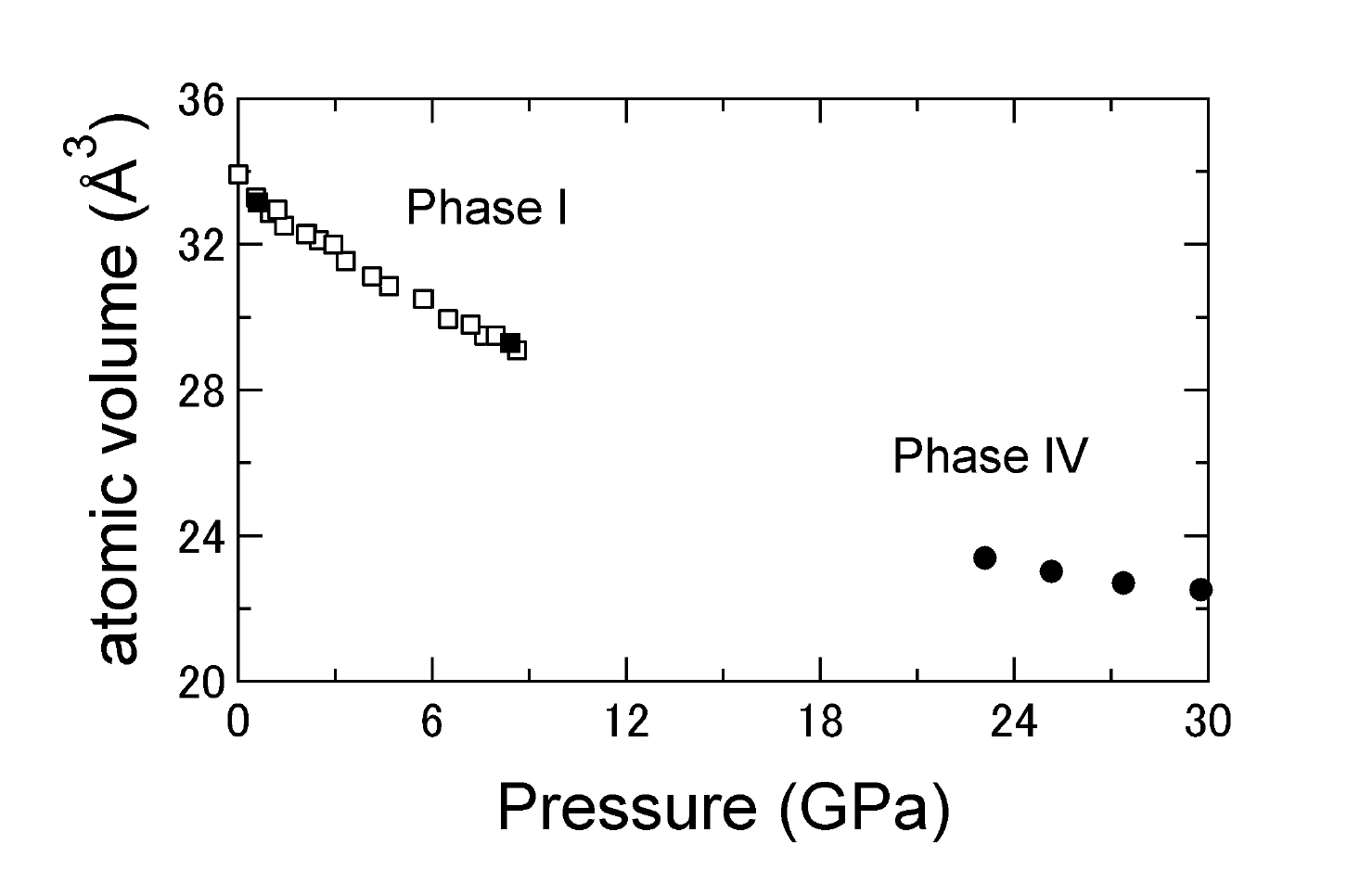}
\caption{\label{fig:G(r) on compression} Pressure dependences of the atomic volumes for phases I and IV. Closed circles indicate the atomic volume of phase IV. Closed and open squares indicate the atomic volume of the present and previous data of phase I, respectively.}
\end{figure}

\begin{figure}
\includegraphics[width=15cm,bb=0 0 501 493]{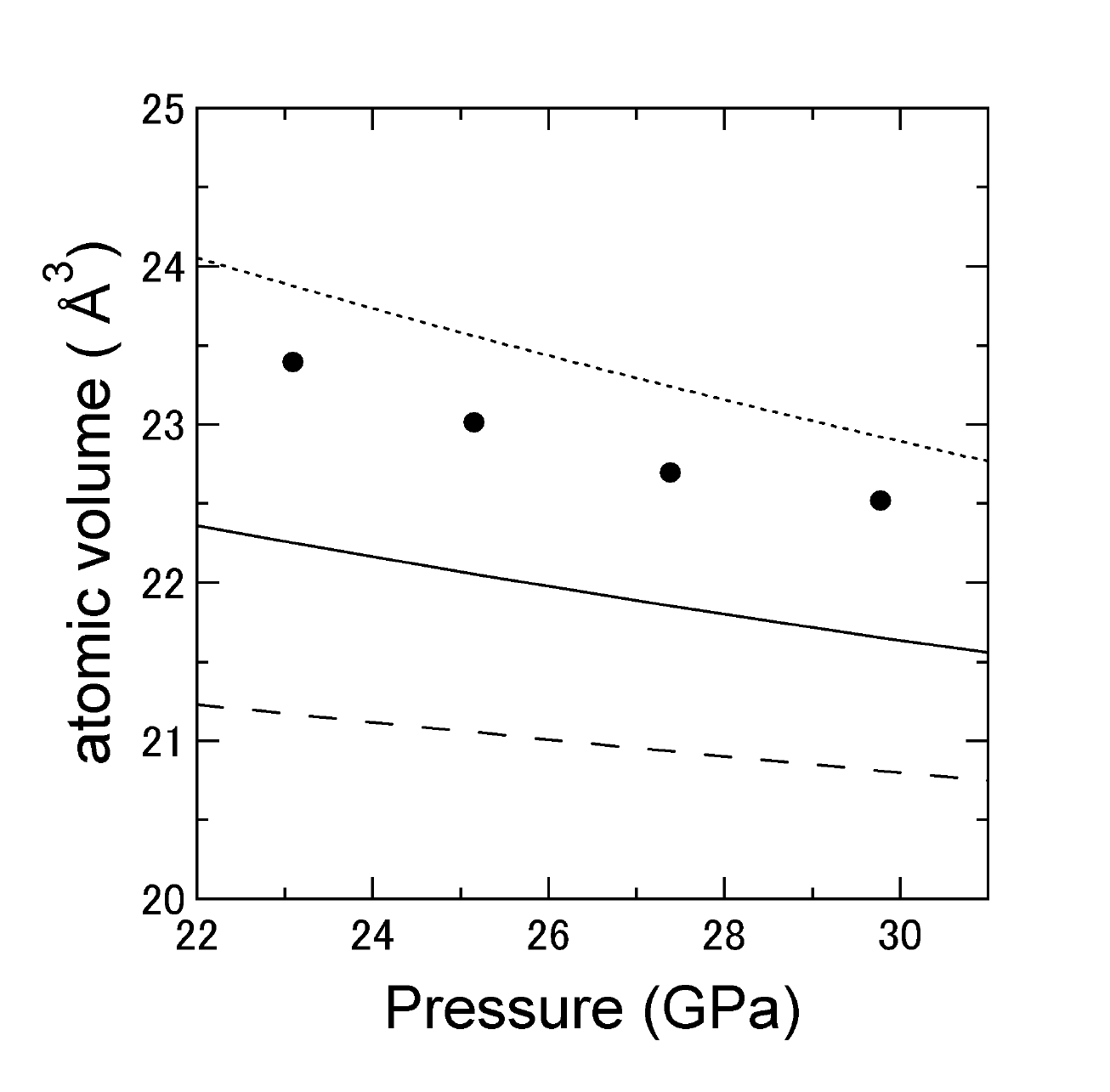}
\caption{\label{fig:Atomic spacing} Pressure dependence of the atomic volumes for high-pressure bcc-phase of Bi$_2$Te$_3$, Bi, and Te. Closed ciecles are experimentally observed Bi$_2$Te$_3$. Solid line estimates Bi$_{0.4}$Te$_{0.6}$ from Vegard's law. Dotted line is from \cite{24} for Bi. Dashed line is from \cite{25} for Te.}
\end{figure}


\begin{references}

\bibitem{1} H. J. Goldsmid, \textit{Thermoelectric Refrigeration} (Plenum Press, New York, 1964).

\bibitem{2} W. M. Yim and F. D. Rosi, Solid-State Electronics \textbf{15}, 1121 (1972).

\bibitem{3} D. M. Rowe, \textit{CRC Handbook of Thermoelectrics} (CRC Press Inc., New York, 1995).

\bibitem{4} Y. Feutelais, B. Legendre, N. Rodier and V. Agafonov, Mat. Res. Bull. \textbf{28}, 591 (1993).

\bibitem{5} C.-Y. Li, A. L. Ruoff, and C. W. Spencer, J. Appl. Phys. \textbf{32}, 1733 (1961).

\bibitem{6} E. S. Itskevich, S. V. Popova and E. Y. Atabaeva, Sov. Phys.-Dokl. \textbf{8}, 1086 (1964).

\bibitem{7} \'{E}. Ya. Atabaeva, E. S. Itskevich, S. A. Mashkov, S. V. Popova, and L. F. Vereshchagin, Sov. Phys.-Solid State \textbf{10}, 43 (1968).

\bibitem{8} L. F. Vereshchagin, E. Y. Atabaeva and N. A. Bendaliani, Sov. Phys.-Solid State \textbf{13}, 2051 (1972).

\bibitem{9} M. A. Il'ina and E. S. Itskevich, Sov. Phys.-Solid State \textbf{13}, 2098 (1972).

\bibitem{10} M. A. Il'ina and E. S. Itskevich, Sov. Phys.-Solid State \textbf{17}, 89 (1975).

\bibitem{11} A. Nakayama, M. Einaga, Y. Tanabe, S. Nakano, F. Ishikawa and Yuh Yamada, High Pressure Research \textbf{29}, 245 (2009).

\bibitem{12} M. Einaga Y. Tanabe, A. Nakayama, A. Ohmura, F. Ishikawa and Yuh Yamada, J. Phys. Conf.: Ser. \textbf{215}, 012036 (2010).

\bibitem{13} C. Zhang, L. Sun, Z. Chen, X. Zhou, Q. Wu, W. Yi, J. Guo, X. Dong and Z. Zhao, Cond-mat. supr-con. arXiv:1009.3746 (2010).

\bibitem{14} J. L. Zhang, S. J. Zhang, H. M. Weng, W. Zhang, L. X. Yang, Q. Q. Liu, S. M. Feng, X. C. Wang, R. C. Yu, L. Z. Cao, L. Wang, W. G. Yang, H. Z. Liu, W. Y. Zhao, S. C. Zhang, X. Dai, Z. Fang and C. Q. Jin, Cond-mat. supr-con. arXiv:1009.3691 (2010).

\bibitem{15} M. K. Jacobsen, R. S. Kumar, A. L. Cornelius, S. V. Sinogeiken and M. F. Nicol, AIP Conf. Proc. \textbf{955}, 171 (2007).

\bibitem{16} L. Fu and C. L. Kane, Phys. Rev. B \textbf{76}, 045302 (2007).

\bibitem{17} Y. L. Chen, J. G. Analytis, J.-H. Chu, Z. K. Liu, S.-K. Mo, X. L. Qi, H. J. Zhang, D. H. Lu, X. Dai, Z. Fang, S. C. Zhang, I. R. Fisher, Z. Hussain and Z.-X. Shen, Science \textbf{325}, 178 (2009).

\bibitem{18} K. Takemura, P. Ch. Sahu, Y. Kunii and Y. Toma, Rev. Sci. Instrum. \textbf{72}, 3873 (2001).

\bibitem{19} C.-S. Zha, H. -K. Mao and R. J. Hemley, proc. Natl. Acad. Sci. \textbf{97}, 13494 (2000).

\bibitem{20} H. M. Rietveld, J. Appl. Crystallogr. \textbf{2}, 65 (1969).

\bibitem{21} F. Izumi and T. Ikeda, Mater. Sci. Forum \textbf{198}, 321 (2000).

\bibitem{22} L. Vegard, Z. Phys. \textbf{5}, 17 (1928).

\bibitem{23} A. R. Denton and N. W. Ashcroft, Phys. Rev. A \textbf{43}, 3161 (1991).

\bibitem{24} Y. Akahama and H. Kawamura, J. Appl. Phys. \textbf{92}, 5892 (2002).

\bibitem{25} G. Parthasarathy and W. B. Holzapfel, Phys. Rev. B, Rap. Comm. \textbf{37}, 8499 (1988).

\bibitem{26} V. M. Glazov, Inorg. Mater. \textbf{20}, 1068 (1984).

\bibitem{27} A. Ohmura (private communication).

\bibitem{28} U. Haussermann, O. Degtyareva, A. S. Mikhaylushkin, K. Soderberg, S. I. Simak, M. I. McMahon, R. J. Nelmes, and R. Norrestam, Phys. Rev. B \textbf{69}, 134203 (2004).

\bibitem{29} J. S. Lannin, Solid State Comm. \textbf{29}, 159 (1979).

\bibitem{30} J. R. Drabble and C. H. L. Goodman, J. Phys. Chem. Solids \textbf{5}, 142 (1958).



\end{references}
\end{document}